\documentclass[12pt]{iopart}
\expandafter\let\csname equation*\endcsname\relax
\expandafter\let\csname endequation*\endcsname\relax
\usepackage{amsmath,amssymb}
\usepackage{cite}
\usepackage{graphicx}

\begin{document}

\title[ ]{Quantum-kinetic perspective on photovoltaic device operation in nanostructure-based solar cells}

\author{Urs Aeberhard}

\address{IEK-5 Photovoltaik, Forschungszentrum J\"ulich, 52425 J\"ulich, Germany}
\ead{u.aeberhard@fz-juelich.de}
\vspace{10pt}

\begin{abstract}
The implementation of a wide range of novel concepts for next-generation high-efficiency solar cells is based on nanostructures with configuration-tunable optoelectronic properties. On the other hand, effective nano-optical light-trapping concepts enable the use of ultra-thin absorber architectures. In both cases, the local density of electronic and optical states deviates strongly from that in a homogeneous bulk material. At the same time, non-local and coherent phenomena like tunneling or ballistic transport become increasingly relevant. As a consequence, the semi-classical, diffusive bulk picture conventionally assumed may no longer be appropriate to describe the physical processes of generation, transport, and recombination governing the photovoltaic operation of such devices. In this review, we provide a quantum-kinetic perspective on photovoltaic device operation that reaches beyond the limits of the standard simulation models for bulk solar cells. Deviations from bulk physics are assessed in ultra-thin film and nanostructure-based solar cell architectures by comparing the predictions of the semi-classical models for key physical quantities such as absorption coefficients, emission spectra, generation and recombination rates as well as potentials, densities and currents with the corresponding properties as given by a more fundamental description based on non-equilibrium quantum statistical mechanics. This advanced approach, while paving the way to a comprehensive quantum theory of photovoltaics, bridges simulations at microscopic material and macroscopic device levels by providing the charge carrier dynamics at the mesoscale.
\end{abstract}

\newpage

\section{Introduction} 
Nanostructures have been considered for photovoltaic applications since the advent of epitaxial growth, with the quantum well solar cell (QWSC) development as one of the pioneering activities \cite{barnham:90}. As early as 2001, the rapidly growing interest in nanostructures and the challenge of efficient as well as cost-effective photovoltaic energy conversion had led to a seminal workshop on nanostructures for photovoltaics in Dresden, where a large number of concepts and structures with different functionalities (optical, electronic, vibrational) were discussed, which for the most part belonged to the then emerging field of third generation photovoltaics \cite{green:01}. This included quantum well and quantum dot heterostructures for tunable absorbers in multi-junction solar cells \cite{green:00}, for the formation of intermediate bands for sequential multi-photon absorption \cite{marti:01_ted}, for the modification of carrier-phonon interaction to slow down cooling in hot-carrier solar cells \cite{conibeer:08}, and for the engineering of efficient carrier multiplication processes in colloidal quantum dots \cite{ellingson:05}. Additionally, a wide range of optical nanostructures have been proposed for nanophotonic or plasmonic light-trapping, spectral splitting and up-conversion or down-shifting \cite{atwater:10,mokkapati:12}. In many cases, efficiencies far beyond the single junction Shockley-Queisser limit were predicted based on detailed balance calculations. Since then, research groups around the globe have been working on the implementation of those promising concepts. However, to date, in very few cases only could efficiencies be reached that are close to the detailed balance limits. The poor performance of nanostructure-based solar cells is mainly attributed to a detrimental increase of recombination losses due to a larger number of internal interfaces that are prone to the accumulation of recombination-active defect states, and to wave function localization in nanostructure states which leads to trapping and slow carrier extraction. On the other hand, the notable exception of the strain-balanced QWSC \cite{adams:10} shows that -- if the aforementioned problems can be circumvented -- a nanostructure-based solar cell holds the potential to outperform the conventional bulk devices. 
The large discrepancy between the predicted and the achieved efficiency values questions the adequacy of the theoretical approach used to obtain these performance estimates and calls for a more realistic assessment of the potential of a given nanostructure-based solar cell concept that is capable of reaching beyond the semi-classical bulk picture \cite{ae:jstqe_13}. In this Review, such a theoretical picture of extended validity is discussed and its predictions are compared against those provided by the conventional semiclassical framework for the modeling of solar cell device characteristics in the case of nanostructure-based architectures.  
The Review is organized as follows. In the section following this introduction, the conventional solar cell device simulation models based on global detailed balance  relations and macroscopic continuity equations with drift-diffusion currents and semiclassical rate terms  are briefly reviewed, after which  the basic formalism of the quantum-kinetic framework underlying the advanced picture is outlined. In a third section, deviations from flat-band bulk behavior are identified for solar cell devices with nanostructure components of decreasing dimensionality, from ultra-thin absorbers to quantum well and quantum dot architectures, by analyzing the results of the different simulation approaches.

\section{Modeling approaches for nanostructure photovoltaics}
The central figure of merit of a solar cell device is its conversion efficiency $\eta=P^{\nearrow}_{\mathrm{el}}/P_{\mathrm{rad}}^{\swarrow}$, where $P^{\nearrow}_{\mathrm{el}}$  is the electrical output power and $P_{\mathrm{rad}}^{\swarrow}$ is the incident radiative power. The main task of any solar cell model therefore consists in providing the electrical output power as a function of the incident photon flux. The electrical power is defined by the product of charge current $I$ and voltage $V$ at a certain point of operation, such that the efficiency is given by $\eta=\max_{V}\{J(V)\cdot V\}/J_{\gamma}$, where $J$  is the charge current density and $J_{\gamma}$  is the illumination intensity. Thus, the key quantity to determine is the current-voltage characteristics  $J(V)$ of the solar cell, which is an integral and scalar quantity, but which contains the full complexity of the device in terms of optical and electronic properties in any of its constituent parts, including nanostructured regions exhibiting peculiar non-bulk-like dynamics. In the following, the evaluation of the current-voltage characteristics is discussed for different levels of complexity in the hierarchy of modeling approaches. The focus thereby is on the electronic aspects rather than optical engineering, as the latter subject is well-covered by the literature.

\subsection{Thermodynamical models for global detailed balance limits}

In the most elementary of the methods, the device characteristics are obtained based on external radiative properties only, by equating the extracted charge current density with the difference of absorbed and emitted photon flux \cite{araujo:94}, 
\begin{align}
J(V)=-q\big\{\Phi_{\mathrm{abs}}(\phi_{0\gamma}^{\swarrow})-\Phi_{\mathrm{em}}(V)\big\},
\end{align}
where $q$  is the elementary charge, $\phi_{0\gamma}^{\swarrow}$ is the incident photon flux and $V$ is the voltage applied between contacts. In the global radiative detailed balance limit, the two quantities are related via their dependence on the absorptance of the device \cite{araujo:94,wuerfel:82},   
\begin{align}
\Phi_{\mathrm{abs}}(\phi_{0\gamma}^{\swarrow})&=\int dE_{\gamma}\,\phi_{0\gamma}^{\swarrow}(E_{\gamma})a(\Omega^{\swarrow},E_{\gamma}),\label{eq:phiabs}\\
 \Phi_{\mathrm{em}}(V)&\approx\int dE_{\gamma}\int d\Omega\,a(\Omega,E_{\gamma})\bar{\phi}_{\mathrm{bb}}(\Omega,E_{\gamma})\big\{\exp(qV/k_{B}T)-1\big\},\label{eq:phiem}
\end{align}
where $E_{\gamma}$ is the photon energy, $a$ is the absorptance (at angle of incidence $\Omega^{\swarrow}$) -- related to the absorption coefficient $\alpha$ and refractive index $n$ via some optical model for the light propagation, such as, e.g., Lambert-Beer’s law or the Transfer-Matrix Method (TMM) for coherent wave propagation -- and $\bar{\phi}_{\mathrm{bb}}$ is the black-body radiation flux for emission into solid angle $\Omega$. In the above formulation, the model assumes perfect carrier extraction due to infinite mobility, corresponding to flat quasi-Fermi levels split by the voltage at the contacts, $\Delta\mu=qV$. The model can be extended to the case of finite mobility and non-radiative recombination by replacing in Eq.~(\ref{eq:phiem}) the absorptance with the external quantum efficiency \cite{rau:07}. However, there are limits in the applicability of this reciprocity relation that stem from the voltage or illumination dependence of the external quantum efficiency which lead to the breakdown of the superposition principle. 

\subsection{Hybrid models for local characteristics from macroscopic parameters}
In order to capture effects of finite mobility, the transport of photogenerated or electronically injected charge carriers needs to be considered. For a detailed investigation of the impact of a specific nanostructured device component on the overall device characteristics, the local charge carrier dynamics has to be included in the model. In general, this is achieved by combining a drift-diffusion model for carrier transport (upper/lower sign is for electrons/holes)
\begin{align} 
\mathbf{J}_{c}[\mu_{c},D_{c}](\mathbf{r})=\mp q \big\{\pm\rho_{c}(\mathbf{r})\mu_{c}(\mathbf{r})\nabla
\phi(\mathbf{r})-D_{c}(\mathbf{r})\nabla\rho_{c}(\mathbf{r})\big\}\qquad (c=e,h)\label{eq:dd_curr}
\end{align}
with Fermi-Golden-Rule rates for the carrier generation ($\mathcal{G}$) and recombination ($\mathcal{R}$)
\begin{align}
\mathcal{G}[\alpha,n](\mathbf{r})&=\int dE_{\gamma}\,\eta_{\mathrm{gen}}(E_{\gamma})\alpha(\mathbf{r},E_{\gamma})\phi_{\gamma}[\alpha,n](\mathbf{r},E_{\gamma}),\label{eq:genrate}\\
\mathcal{R}[\alpha,n](\mathbf{r})&=\mathcal{B}(\mathbf{r})\rho_{e}(\mathbf{r})\rho_{h}(\mathbf{r}),\quad \mathcal{B}(\mathbf{r})=n_{i}^{-2}\int dE_{\gamma}\,\alpha(\mathbf{r},E_{\gamma})\tilde{\phi}_{bb}[n](E_{\gamma}),\label{eq:recrate}
\end{align}
obtained from the local microscopic electronic structure information, in a balance equation for charge continuity
\begin{align}
\mp q^{-1}\nabla \cdot \mathbf{J}_{c}[\mu_{c},D_{c}]=\mathcal{G}[\alpha,n]-\mathcal{R}[\tau]\label{eq:baleq}
\end{align}
coupled to the Poisson equation for the electrostatic potential $\phi$,
\begin{align}
\epsilon_{0}\nabla\cdot\left\{\varepsilon(\mathbf{r})\nabla
\phi(\mathbf{r})\right\}=q\left\{\rho_{e}(\mathbf{r})-\rho_{h}(\mathbf{r})-N_{\textrm{dop}}(\mathbf{r})\right\}.\label{eq:poisseq}
\end{align}
In the above equations,  $\mu$ is the mobility, $D$ the diffusion constant, $\rho$ ($n_{i}$) the (intrinsic) carrier density,  $\phi_{\gamma}$ is the local photon flux due to the external illumination,  $\eta_{\mathrm{gen}}$ is the fraction of photons generating electron-hole pairs, $\tilde{\phi}_{\mathrm{bb}}[n]$  is the angle-integrated black-body flux for isotropic emission into medium with refractive index $n$, $\tau$ is the carrier lifetime associated with the recombination process, $\varepsilon_{0}$ and $\varepsilon$ are the free space and relative permittivities, and $N_{\mathrm{dop}}$ is the density of ionized dopants. Conventionally, the carrier density is expressed in terms of an effective density of states $\mathcal{N}$ that reflects the electronic structure close to the band edge, and of the carrier distribution function, for which it is common to use Boltzmann statistics with quasi-Fermi level $E_{F_{c}}$: 
\begin{align}
\rho_{c}(\mathbf{r})=\mathcal{N}_{c}(\mathbf{r})\exp\{[\pm E_{F_{c}}(\mathbf{r})\mp E_{B}(\mathbf{r})]/k_{B}T\},\label{eq:sc_dens}
\end{align}
where upper (lower) sign applies to electrons (holes), and $E_{B}$ is the band edge energy. Solution of Eqns. \eqref{eq:dd_curr}, \eqref{eq:baleq}-\eqref{eq:sc_dens} provides the current-voltage characteristics as a function of quasi-Fermi-levels $E_{F_{n,p}}$ for electrons and holes and of the electrostatic potential $\phi$. Modified versions of Expr.~(\ref{eq:baleq}) have been used in cases with vanishing current between absorbers and/or absent coupling to contact states \cite{ae:jstqe_13}.
Such a hybrid approach is limited in validity by the assumptions underlying the drift-diffusion picture, i.e., band-like transport with completely thermalized carrier distributions, which does not include any quantum effects such as confinement, tunneling or ballistic transport on very short length scales. Moreover, in most cases, the model used for the electronic structure relies on the flat band bulk picture, which is not applicable in nanostructure regions. 

\subsection{Quantum kinetic models for microscopic non-equilibrium dynamics}
The challenges of describing opto-electronic device operation under consideration of quantum effects are manifold. Firstly, one has to treat an open quantum system, which in principle requires a description based on scattering states rather than the eigenstates provided by the solution of Schr\"odinger's equation for the closed system. Due to the essential inclusion of light-matter interaction and the sizable effects of electron-phonon coupling under the standard condition of room temperature operation, a mixed state representation is indicated. Among the suitable theories, the non-equilibrium Green’s function (NEGF) formalism is most versatile and powerful and has found wide-spread application in the modeling of nanostructure-based quantum opto-electronic devices such as photodetectors based on QW \cite{henrickson:02} and QD \cite{naser:07}, QW lasers \cite{pereira:98}, quantum cascade lasers \cite{lee:prb_02,kubis:09}, and QW LEDs \cite{steiger:iwce_09}. In the field of nanostructure photovoltaics, applications of the NEGF formalism so far include carbon nanotube photodiodes \cite{stewart:05}, multi-QW and QW superlattice solar cells \cite{ae:prb_08, ae:nrl_11}, nanowire solar cells \cite{buin:13}, QD superlattice solar cells \cite{ae:oqel_12,berbezier:15}, ultra-thin absorber devices \cite{cavassilas:15,ae:jpv_16}, and QW tunnel junctions for multi-junction solar cells \cite{ae:prb87_13}. 
For a detailed introduction to the NEGF approach for the simulation of nanostructure-based solar cell devices, the reader is referred to Ref.~\cite{ae:jcel_11}. Here, we give only the elements that are linked to the device characteristics and which are required for relation to the other approaches introduced above. In the NEGF picture, a microscopic conservation law similar to the continuity equation \eqref{eq:baleq} can be formulated,
\begin{align}
 \nabla \cdot \mathbf{J}[G]=\mathcal{G}[G,\Sigma]-\mathcal{R}[G,\Sigma],	\label{eq:negf_baleq}
\end{align}
where now both the (electron) charge current on the left hand side,
\begin{align}
\mathbf{J}[G](\mathbf{r})=\lim_{\mathbf{r}'\rightarrow
	\mathbf{r}}\frac{e\hbar}{m_{0}}\big(\nabla_{\mathbf{r}}-\nabla_{\mathbf{r}'}\big)
\int\frac{dE}{2\pi}G^{<}(\mathbf{r},\mathbf{r}',E),
\end{align}
and the rate expressions on the right hand side,
\begin{align}
\mathcal{G}/\mathcal{R}[G,\Sigma](\mathbf{r})=\int d^{3}r'\int dE\,G^{\gtrless}(\mathbf{r},\mathbf{r}',E)\Sigma^{\lessgtr}[G^{\lessgtr}](\mathbf{r'},\mathbf{r},E),\label{eq:negf_rates}
\end{align}
are formulated in terms of the charge carrier Green’s functions $G$, and of self-energies  $\Sigma$ encoding the interaction of the carriers with the environment in terms of scattering and coupling to contacts.  The Green’s functions follow from the steady-state Dyson and Keldysh equations in the framework of non-equilibrium quantum statistical mechanics \cite{kadanoff:62,keldysh:65},
\begin{align}
G^{R(A)}({\mathbf r_{1}},{\mathbf
	r}_{1'},E)&=G_{0}^{R(A)}({\mathbf r_{1}},{\mathbf
	r}_{1'},E)+\int d^{3}r_{2}\int
d^{3}r_{3}G_{0}^{R(A)}({\mathbf r}_{1},
{\mathbf r}_{2},E)\nonumber\\&\quad\quad\times\Sigma^{R(A)}({\mathbf r}_{2},{\mathbf
	r}_{3},E) G^{R(A)}({\mathbf r}_{3},{\mathbf
	r}_{1'},E),\\
 G^{\lessgtr}({\mathbf r}_{1},{\mathbf
	r}_{1'},E)=&\int d^{3}r_{2}\int d^{3}r_{3} 
G^{R}({\mathbf
	r}_{1}, {\mathbf r}_{2},E)\Sigma^{\lessgtr}({\mathbf
	r}_{2},{\mathbf r}_{3},E) G^{A}({\mathbf
	r}_{3},{\mathbf r}_{1'},E).\label{eq:keldysh}
\end{align}
The interaction component of the carrier self-energy  $\Sigma$ encodes the coupling of electrons and holes to photons and phonons, enabling the essential description of photogeneration, radiative recombination and relaxation of carriers.  The interaction of charge carriers with electromagnetic radiation is described on two levels: for the coherent processes of absorption and stimulated emission, minimal coupling to the classical vector potential $\mathbf{A}$ - obtained, e.g., by solving Maxwell's equations - in dipole approximation provides the self-energy
\begin{align}
\Sigma^{\lessgtr}(\mathbf{r},\mathbf{r}',E)=\Big(\frac{e}{m_{0}}\Big)^{2}\sum_{\mu\nu}\int
dE_{\gamma} \Big[&A_{\mu}(\mathbf{r},E_{\gamma})
p_{cv}^{\mu}(\mathbf{r}) G^{\lessgtr}(\mathbf{r},\mathbf{r}',E\mp
E_{\gamma})\nonumber\\&\times A_{\nu}^{*}(\mathbf{r}',E_{\gamma})p_{cv}^{\nu*}(\mathbf{r}')\Big].\label{eq:se_phot_coh}
\end{align}
Here, greek subscripts denote polarization indices, and $\mathbf{p}_{cv}$ is the momentum matrix. Spontaneous emission, on the other hand, requires incoherent coupling to the entirety of photon modes available as encoded in the photon Green’s function ${\mathcal{D}}$. The corresponding self-energy on the level of the self-consistent Born approximation (SCBA) from second-order perturbation theory in the interaction of charge carriers with the quantized photon field reads
\begin{align}
\Sigma^{e\gamma,\lessgtr}(\mathbf{r},\mathbf{r}',E)=i\hbar\mu_{0}\Big(\frac{e}{m_{0}}\Big)^2&\sum_{\mu\nu}\lim_{\mathbf{r}''\rightarrow\mathbf{r}} \Big[\frac{1}{2}\left\{\hat{p}^{\mu}(\mathbf{r})-\hat{p}^{\mu}(\mathbf{r}'')\right\}\hat{p}^{\nu}(\mathbf{r}')
\nonumber\\&\times \int\frac{dE'}{2\pi\hbar}\mathcal{D}^{\lessgtr}_{\mu\nu}(\mathbf{r}'',\mathbf{r'},E') G^{\lessgtr}(\mathbf{r},\mathbf{r'},E-E')\Big],\label{eq:se_phot_inc}
\end{align}
where $\hat{\mathbf{p}}=-i\hbar\nabla$ is the momentum operator. A similar SCBA self-energy is used for the description of electron-phonon interaction. The contact self-energies, on the other hand, implement the open boundary conditions required for the description of charge carrier extraction and injection, and have the general form $\Sigma^{B}=\mathcal{T}g^{B}\mathcal{T}^{\dagger}$ , where $\mathcal{T}$ encodes the coupling to the contact and $g^{B}$ is the surface Green’s function of the electrode. The level of electronic injection is set by the chemical potentials of left and right electrodes, with their separation corresponding to the applied bias voltage, $\mu_{R}-\mu_{L}=~qV$. The evaluation of the Green’s functions (GF) is again self-consistently coupled to the computation of the electrostatic potential via Poisson’s equation (8) through the expression of the carrier density in terms of the GF:
\begin{align}
\rho_{e/h}(\mathbf{r})=\mp i\int
\frac{dE}{2\pi}G_{c/v}^{\lessgtr}(\mathbf{r},\mathbf{r},E).\label{eq:negf_dens}
\end{align}
In contrast to the semiclassical bulk picture, this formalism provides non-local microscopic non-equilibrium dynamics in arbitrary potentials under explicit consideration of the relevant scattering processes, which covers transport from ballistic to diffusive regimes and rigorous treatment of tunneling and size quantization effects. Due to the consistent description of spectral quantities and integral characteristics, the approach is ideally suited to mediate between microscopic material properties and nanostructure configurations on the one hand, and macroscopic device behavior on the other hand, providing a natural framework for multi-scale simulation of nanostructure-based solar cells. 
On the other hand,  a multi-scale approach is also indicated due to the very large computational cost of the NEGF approach, that stems from the microscopic resolution in energy, momentum and real space, and of the requirement for self-consistent computations, and which currently limits the applicability of the formalism to nanostructure regions of mesoscopic extension. Massively parallel implementations of the approach may help to cope with the immense computational load, but are challenging due to the communication overhead resulting from the coupling in energy and momentum space when considering inelastic scattering, and non-locality in real space requiring the computation of off-diagonal GF elements.

\section{Applications}
Two common aspects of nano-scale or nanostructure absorbers are in the focus of the investigation here: the effects of departure from flat band potentials due to strong built-in fields, and the impact of finite size effects such as the presence of non-classical contact regions. These are investigated by application of the   simulation approaches introduced above to three different generic solar cell device architectures which deviate increasingly from the bulk situation in terms of the spatial confinement of the absorber states: ultra-thin absorber solar cells, single quantum well photo-diodes and quantum dot superlattice solar cells.  

\subsection{Ultra-thin solar cells}
\begin{figure}[t]
	\begin{center}
		\includegraphics[width=1\textwidth]{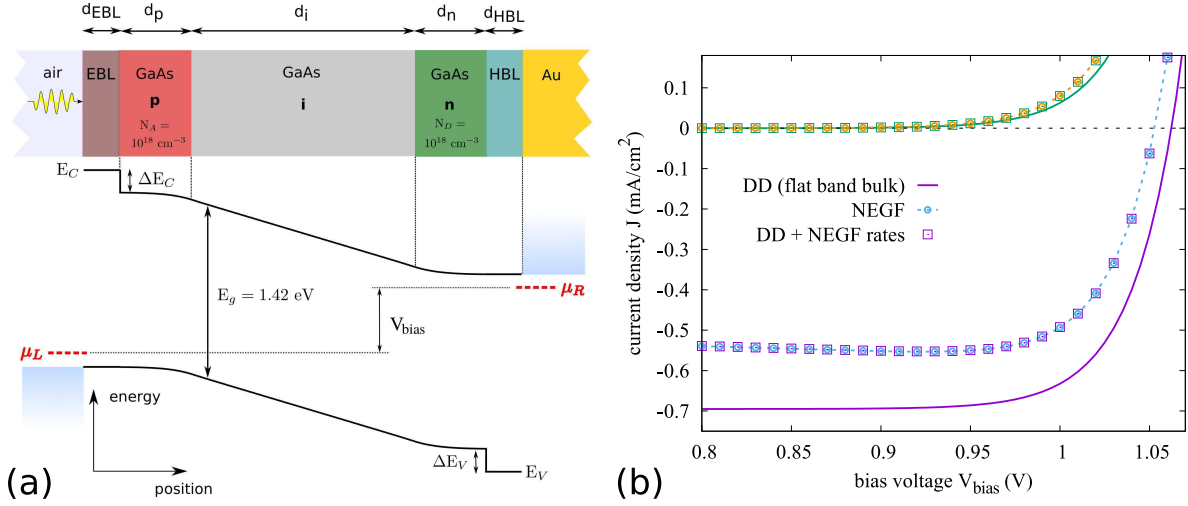}
\caption{(a) Layer structure, doping levels and corresponding band profile of an ultra-thin GaAs $p$-$i$-$n$ solar cell with carrier selective contacts \cite{ae:apl_16}. (b) Device characteristics of the solar cell device in (a) in the dark and under monochromatic illumination with photon energy $E_{\gamma}=1.44$ eV and intensity of 0.1 kW/m$^{2}$. While there is a significant deviation of the NEGF result from the drift-diffusion characteristics based on flat band bulk material parameters, use of the NEGF rates in the drift-diffusion formalism reproduces the full NEGF result, pointing at ideal carrier extraction \cite{ae:jpv_16}.\label{fig:fig_1}}
	\end{center}
\end{figure}

\begin{figure}[!h]
	\begin{center}
		\includegraphics{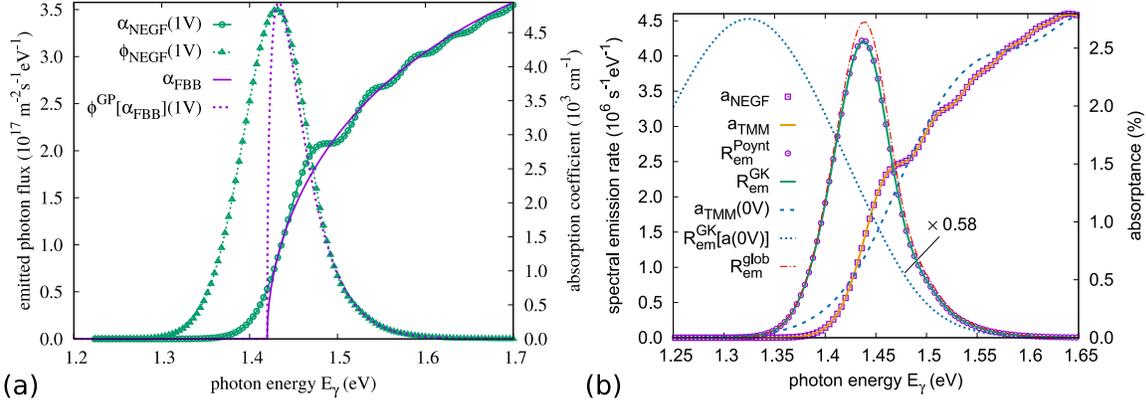}

	\caption{(a) Absorption coefficient and emission rate at the center of the intrinsic region, as provided by a flat-potential parabolic two-band bulk model (lines) using the Van Roosbroeck- Shockley formalism, and as obtained from the NEGF formalism (symbols) for the same two-band model, but including the actual band profile \cite{ae:jpv_16}. (b) Absorptance and normal emission at the left surface of the solar cell, demonstrating the breakdown of the photovoltaic reciprocity between external quantum efficiency at zero bias voltage and luminescent emission at large forward bias: the strong modification of the absorptance with bias invalidates the use of the short circuit absorptance in the Generalized Kirchhoff law for the emission. The difference in the emission with respect to the dash-dotted line corresponding to the integrated internal emission rate is a signature of photon recycling effects \cite{ae:17_prl}.\label{fig:fig_2}}
		\end{center}
\end{figure}

Recently, strongly enhanced to almost complete light absorption in ultra-scaled absorber layers with thicknesses in the deep sub-wavelength regime and nanophotonic light trapping has been shown both theoretically \cite{wang:13,llorens:14} and experimentally \cite{massiot:12,massiot:13,massiot:14}.  This opens up a route to highly efficient photovoltaic devices with vastly reduced material consumption and increased radiation tolerance \cite{wang:13_jpv,yang:14_jap,vandamme:15,hirst:16}. The understanding and design of quasi-confined or resonant optical modes is key to the realization of ultra-thin solar cells, however, the strong reduction of spatial extension affects also the electronic properties. While there is no heterostructure potential in the absorber region, the non-classical contact regions occupy a significant fraction of the device and have a pronounced influence on the device characteristics. Figure \ref{fig:fig_1}(a) displays the structure of the ultra-thin GaAs $p$-$i$-$n$ device under consideration here. The total thickness of the device  is less than 100 nm. The light is incident from the left without any anti-reflection coating, and is reflected at a gold mirror forming the right contact. The gold layer is considered explicitly only in the optical simulation.  The optical model is based on the transfer-matrix method in all cases, thus including the effects of multiple reflections. The optical rates and transport equations are computed either using the  semiclassical formalism [Eqs.~\eqref{eq:dd_curr}-\eqref{eq:sc_dens}] as implemented in the 1D thin-film solar cell simulator {\it ASA} \cite{pieters:06}, or the quantum-kinetic formalism [Eqs. \eqref{eq:negf_baleq}-\eqref{eq:keldysh}] for a two band effective mass Hamiltonian. In the latter case, while the generation and recombination processes are enabled directly via the single particle interaction self-energies \eqref{eq:se_phot_coh} and \eqref{eq:se_phot_inc} for coherent and incoherent coupling to the (quantized) electromagnetic field, respectively, the representation of the optical rates as in \eqref{eq:negf_rates} can be used to express the absorption coefficient in terms of microscopic quantities:
\begin{align}
\alpha(z,E_{\gamma})\approx&\frac{\hbar c_{0}}{6
	n_{r}(z,E_{\gamma})E_{\gamma}}\sum_{\mu}\int dz'
\mathrm{Re}\Big[i\hat{\Pi}_{\mu\mu}(\mathbf{0},z',z,E_{\gamma})\Big],
\end{align}
where
\begin{align}
\Pi_{\mu\nu}^{\lessgtr}(\mathbf{q}_{\parallel},z,z',E_{\gamma})=&-i\hbar\mu_{0}\Big(\frac{e}{m_{0}}\Big)^{2}p_{cv}^{\mu*}(z)p_{cv}^{\nu}(z')\mathcal{P}_{cv}^{\lessgtr}(\mathbf{q}_{\parallel},z,z',E_{\gamma})
\end{align}
is the photon self-energy for the electron-hole polarization function
\begin{align}
\mathcal{P}_{cv}^{\lessgtr}(\mathbf{q}_{\parallel},z,z',E_{\gamma})=&\mathcal{A}^{-1}\sum_{\mathbf{k}_{\parallel}}\int
\frac{dE}{2\pi\hbar}G_{c}^{\lessgtr}(\mathbf{k}_{\parallel},z,z',E)G_{v}^{\gtrless}(\mathbf{k}_{\parallel} -\mathbf{q}_{\parallel},z',z,E-E_{\gamma}).\label{eq:polfun}
\end{align}
In the above equations, the slab representation adequate for planar geometries is used, with $z$  denoting the perpendicular coordinate and $\mathbf{k}_{\parallel}$ ($\mathbf{q}_{\parallel}$) the transverse momentum of electrons (photons) associated with the periodic in-plane dimensions. $\mu$ and $\nu$ are polarization indices,  $\mu_{0}$ is the vacuum permeability, $\mathbf{p}_{cv}$ is the interband momentum matrix element and $\mathcal{A}$ is the transverse cross section. In terms of the photon self-energy and Green’s function, i.e., in a picture that allows for a consistent description of absorption and emission, the absorptance appearing in the global detailed balance limit acquires the form ($\hat{\mathcal{D}}\equiv \mathcal{D}^{>}-\mathcal{D}^{<}$, $\hat{\Pi}\equiv \Pi^{>}-\Pi^{<}$)
\begin{align}
a_{\mu\nu}(\mathbf{q}_{\parallel},E_{\gamma})=&-\int dz\int dz'~\Big[\hat{\mathcal{D}}_{v,\mu\nu}(\mathbf{q}_{\parallel},z,z',E_{\gamma})\hat{\Pi}_{\nu\mu}(\mathbf{q}_{\parallel},z',z,E_{\gamma})\Big].\label{eq:negf_absorpt}
\end{align}
In the same picture, the spectral emission rate corresponding to the integrand in the semiclassical expression \eqref{eq:recrate} reads
\begin{align}
r^{\mu}(\mathbf{q}_{\parallel},z,E_{\gamma})=&\sum_{\nu}\int dz'~(2\pi\hbar)^{-1}\mathcal{D}^{>}_{\mu\nu}(\mathbf{q}_{\parallel},z,z',E_{\gamma})\Pi_{\nu\mu}^{<}(\mathbf{q}_{\parallel},z',z,E_{\gamma}).
\end{align}
In the band profile displayed in Fig. \ref{fig:fig_1}(a), two salient features are the sizable bend bending due to the large built-in field, and the carrier selective contacts imposed by blocking layers with potential barriers for minority carriers. Fig. \ref{fig:fig_1}(b) displays the current-voltage characteristics for the case of perfect selectivity ($\Delta E_{C,V}\rightarrow \infty$), in the dark and under  monochromatic  illumination at  $E_{\gamma}=1.44$ eV and an intensity of 0.1 kW/m$^{2}$. The sizable discrepancies in dark and photocurrent between the semiclassical drift-diffusion results and the NEGF characteristics are largely explained by the strong field effects on absorption and emission that lead to deviations from flat band bulk (FBB) properties regarding generation and recombination, as shown in Fig. \ref{fig:fig_2}(a) for the  absorption coefficient and the emission rate at the center of the intrinsic region. In the FBB case, the following expression is used for the absorption coefficient:
\begin{align}
\alpha_{\mathrm{bulk}}^{pb}(E_{\gamma})=&\frac{(2m_{r}^{*})^{\frac{3}{2}}}{2\pi\hbar^2\sqrt{\varepsilon_{b}}\varepsilon_{0}c_{0}E_{\gamma}}\Big(\frac{e}{m_{0}} p_{cv}\Big)^2\sqrt{E_{\gamma}-E_{g}}\Theta(E_{\gamma}-E_{g}),
\end{align} 
with $m_{r}^{*}$ the reduced effective mass, $\varepsilon_{b}$ the background dielectric constant and $E_{g}$   the band gap energy. Remarkably, if the NEGF rates \eqref{eq:negf_rates} are used in the semiclassical balance equation \eqref{eq:baleq}, the NEGF characteristics are reproduced. While the actual transport regime might still be very different in the two cases (ballistic extraction in the NEGF picture versus diffusive band-like transport in the drift-diffusion picture), the coincidence roots in perfect carrier extraction due to the long radiative lifetime \cite{ae:jpv_16}. 

As a direct consequence of the pronounced field effects, the large variation of the longitudinal field with terminal voltage $V$ causes the breakdown of the superposition principle and, eventually, of the photovoltaic reciprocity relation \cite{rau:07} between photocurrent extraction -- expressed by the external quantum efficiency -- and the luminescent emission due to carrier injection under applied bias voltage. Figure \ref{fig:fig_2}(b) shows the change of the absorptance when going from short circuit conditions ($V=0$ V) to large forward bias close to open circuit conditions ($V=1.1$ V). The absorptance  obtained from the Green’s functions by evaluation of Eq. \eqref{eq:negf_absorpt} is in excellent agreement with the quantity  directly obtained form the transfer-matrix model, which validates the photon Green’s function approach. As demonstrated further in Fig. \ref{fig:fig_2}(b), the generalized Kirchhoff law for the emission spectrum based on the absorptance -- i.e., Eq. \eqref{eq:phiem} -- applies only, if the absorptance at the point of operation is used, while the emission based on the absorptance at $V=0$ V exhibits a strong broadening and red-shift. The difference between internal and external emission caused by reabsorption becomes visible in Fig. \ref{fig:fig_2}(b), where the integrated (normal) emission rate is shown ($R^{\mathrm{glob}}_{\mathrm{em}}$, dash-dotted line) together with the (normal) emission from the left surface \cite{ae:17_prl}. 

\begin{figure}[t]
		\begin{center}
		\includegraphics{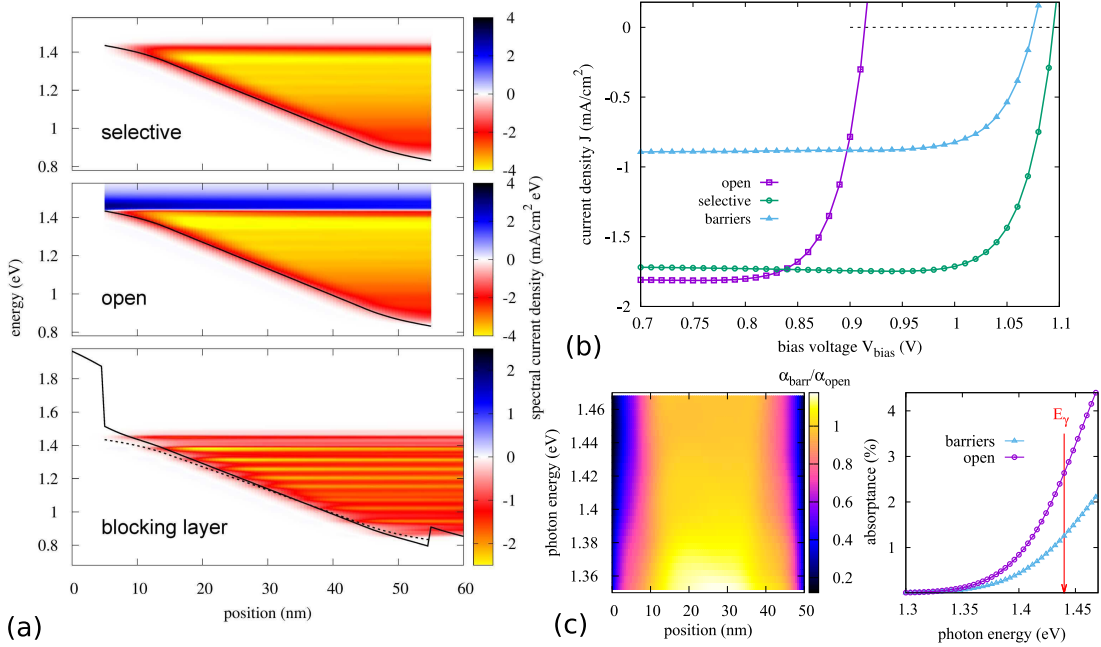}
	\caption{(a) Spectral current flow in a 50-nm thick $p$-$i$-$n$ photodiode with perfectly selective, open and blocking layer contacts, respectively. The open contact configuration leads to pronounced leakage losses at forward bias, while blocking layers at both contacts induce size quantization effects. (b) Current-voltage characteristics for the different configurations: for open contacts, leakage losses result in strong reduction of the open circuit voltage, while contact barriers lead to a reduction in short circuit current. (c) The short circuit current losses induced by barriers are the consequence of reduced absorption due to the appearance of non-classical regions of reduced density of states in the vicinity of the barriers \cite{ae:apl_16}. \label{fig:fig_3}}
		\end{center}
\end{figure}

The second feature that is characteristic for nanoscale absorbers concerns finite size effects related to contact regions that span a non-negligible fraction of the active device and which have a measurable impact on the overall device performance \cite{ae:apl_16}.   Figure \ref{fig:fig_3}(a) displays the spectral current flow in a 50-nm ultrathin device at 0.87 V of forward bias voltage under monochromatic illumination with  $E_{\gamma}=1.44$ eV and at  an intensity of 0.1 kW/m$^{2}$. The three different contact configurations considered are: ideally “selective” ($\Delta E_{C,V}\rightarrow \infty$), “open” ($\Delta E_{C,V}\rightarrow 0$) and with electron and hole “blocking layers” (EBL/HBL) inducing finite potential barriers at the contacts ($0 < \Delta E_{C,V} < \infty$). In the case of GaAs, suitable contact layers can be engineered using alloys such as AlGaAs \cite{vandamme:15} or InGaP \cite{yang:14_jap} or a combination thereof \cite{cavassilas:15}. The ideally selective contact blocks reverse current flow --  i.e.,   leakage current -- of minority carriers completely, while majority carriers are extracted without barrier. In the absence of the barrier for minority carriers (open contact), carriers that are injected at the majority carrier contact or which are photogenerated inside the intrinsic region can leave the device via the minority carrier contact, which represents a significant leakage loss. In the case of photocurrent leakage, the short circuit current is reduced, while leakage of electrically injected carriers contributes to the dark saturation current and hence results in a reduction of open circuit voltage V$_{\textrm{OC}}$. Figure \ref{fig:fig_3}(b) displays the current-voltage characteristics of the different contact configurations. While the loss in V$_{\textrm{OC}}$ is explained by the dark current due to leakage as discussed above, the reduction of short circuit current J$_{\textrm{SC}}$ in the presence of the blocking layers is due to the appearance of non-classical regions of reduced density of states in close proximity of the barriers, which translates to a reduced absorption coefficient [Fig. \ref{fig:fig_3}(c)]. In the case of simultaneous presence of electron and hole blocking layers with finite barriers also for majority carriers, size quantization enhances this effect and leads to both a resonance structure in the spectral current and a strongly reduced absorptance at photon energies close to the band gap, as shown in Fig. \ref{fig:fig_3}(c).

\subsection{Quantum well solar cells}
Quantum well solar cells have been introduced as tunable band gap absorbers in single and multijunction configurations \cite{barnham:90,ned:01}. Advanced strain-balancing approaches allow for the exploitation of a large material parameter space while keeping the density of structural defects low \cite{ned:99_2}. In quantum well solar cells, the absorber contains regions of lower dimensional electronic states that are partially confined in transport direction owing to the variation in band gap and electron affinity of the constituent bulk materials. In conventional multi-quantum-well architectures, the quantum well states are not directly coupled to the contacts, and carriers generated in such localized states have to be transferred to extended continuum states prior to extraction \cite{ae:solmat_10,ae:spie_10}. The escape of carriers proceeds via thermionic emission, direct or phonon assisted tunneling, depending on the temperature, injection level and strength of the built-in field. The situation is different in superlattices of strongly coupled QW \cite{wang:12}, where extraction proceeds via confined states related to miniband formation, with transport regimes ranging from ballistic escape to sequential tunneling assisted by phonon emission \cite{ae:nrl_11,ae:jpe_14}. 

\begin{figure}[b]
	\begin{center}
		\includegraphics{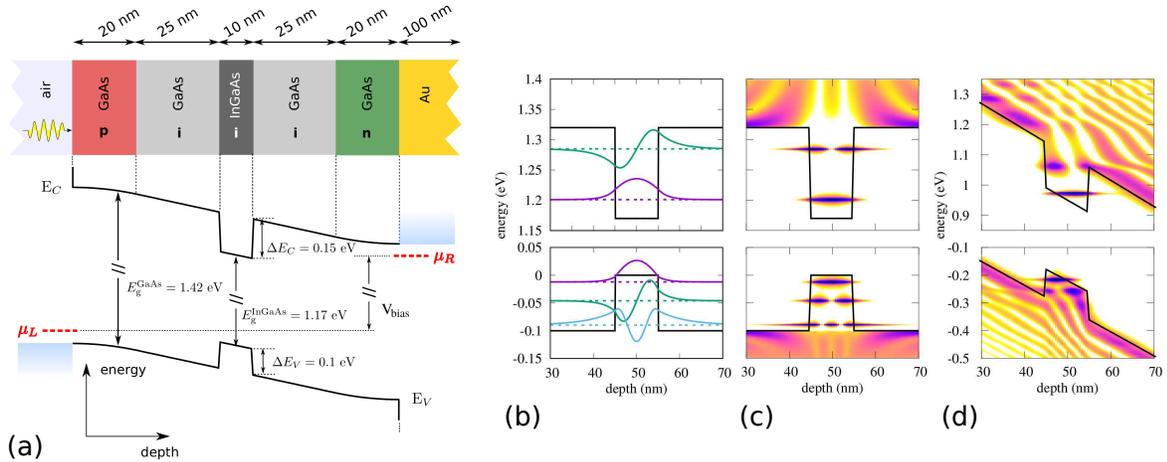}
		
		\caption{ (a) Structure and band profile of a GaAs $p$-$i$-$n$ photodiode with a single InGaAs quantum well embedded in the intrinsic region. (b) Confined states in the QW for flat band potential, as provided by the solution of the effective mass Schr\"odinger equation with closed system boundary conditions. (c) Corresponding local density of states obtained from the NEGF formalism. (d) LDOS for the self-consistent potential at forward bias of 0.8 V, i.e., close to the maximum power point. The deviation from flat band potential leads to Stark shifts and partial unbinding of the QW states.\label{fig:fig_4}}
	\end{center}
\end{figure}

\begin{figure}[!h]
		\begin{center}
		\includegraphics{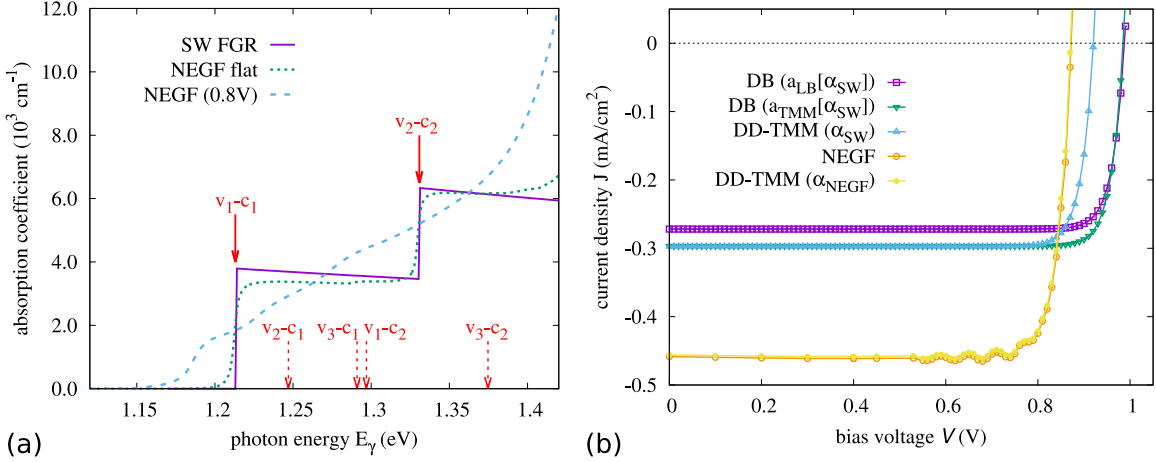}

	\caption{(a) Absorption coefficient of the single quantum well photodiode in the energy range below the bulk band gap of the host material, for the three cases displayed in Fig. 4: from Fermi-Golden-Rule rate (FGR) using the flat band square-well potential states (SW), from NEGF for flat band, and from NEGF for the self-consistent potential at 0.8 V. While the NEGF absorption coefficient reproduces the FGR result up to some broadening effects, including the symmetry-induced selection rules, the absorption coefficient at 0.8 V exhibits a pronounced red-shifting, relaxation of selection rules due to symmetry breaking and deviation from purely two-dimensional density of states. (b) Current-voltage characteristics of the single quantum well photodiode for monochromatic illumination with $E_{\gamma}=1.3$ eV and at an intensity of 0.1 kW/m$^{2}$, as provided by: the detailed balance picture (DB) with absorptance from Lambert-Beer law (LB) or the transfer-matrix method (TMM) for square well absorption coefficient ($\alpha_{SW}$); the semiclassical drift-diffusion-Poisson model (DD) coupled to TMM; and the full NEGF-Poisson model. While the LB and TMM laws provide slightly different absorptance levels, consideration of the absorption coefficient as provided by the NEGF formalism results in semiclassical characteristics that match closely those of the full NEGF model.\label{fig:fig_5}}
		\end{center}
\end{figure}

Here, in order to investigate the impact of the additional deviation from an ultra-thin, but otherwise homogeneous bulk-like absorber, we study a GaAs $p$-$i$-$n$ photodiode with a single InGaAs quantum well embedded in the intrinsic region, as displayed in Fig. \ref{fig:fig_4}(a). As in the case of the ultrathin absorber device, the light is incident from the left, and the gold reflector is considered in the optical simulation only. In Fig. \ref{fig:fig_4}(b), the envelope functions of confined electron and hole states of the field-free quantum well as obtained by solving the effective mass Schr\"odinger equation with closed-system (Dirichlet) boundary conditions are shown. The corresponding local density of states provided by the NEGF formalism based on the same effective mass  Hamiltonian at flat band conditions is displayed in Fig.  \ref{fig:fig_4}(c). This time, the LDOS includes the perturbed continuum above the well, featuring quasi-continuum states that give rise to transmission resonances. While the flat band NEGF picture coincides with the closed-system Schr\"odinger picture for the confined states in terms of resonance energy and spatial variation, the NEGF LDOS for the self-consistent potential at the operating point of 0.8 V of forward bias voltage exhibits strong field effects in terms of Stark shifts and partial unbinding of the higher confinement levels [Fig. \ref{fig:fig_4}(d); please note the difference in energy scale with respect to (b) and (c)]. Following the derivation of the states in the square well potential (SW) under vanishing and self-consistent field, a similar analysis is performed for the absorption coefficient: Figure \ref{fig:fig_5}(a) displays the textbook absorption coefficient as obtained from the Fermi-Golden-Rule (FGR) rate based on the flat band square well potential states,
\begin{align}
\alpha_{\mathrm{SW}}(E_{\gamma})=
f\cdot
E_{\gamma}^{-1}\sum_{i,j}|M_{c_{i}v_{j}}|^2\Theta\big(E_{\gamma}-\varepsilon_{c_{i}v_{j}}\big)
\end{align}
with $\Theta$  the unit step function and
\begin{align}
f=\frac{e^2 m_{r}^{*}P_{cv}^2}{m_{0}^2\hbar n_{r}c_{0}\varepsilon_{0}L_{w}}
\end{align}
where  $m_{r}^{*}$  is the reduced effective mass,  $P_{cv}=\sqrt{E_{P}m_{0}/6}$ is the bulk momentum matrix element for Kane energy  $E_{P}=26.9$ eV, $n_{r}=3.7$ is the refractive index of the bulk material, and $c_{0}$ and $\varepsilon_{0}$ are speed of light and permittivity in vacuum, respectively. The overlap matrix elements  
\begin{align}
M_{c_{i}v_{j}}=\int_{L} dz~ \psi^{*}_{c_{i}}(z)\psi_{v_{j}}(z),
\end{align}
where $L(>L_{w})$  is the normalization length of the envelope functions in $z$-direction $\psi$, introduce the optical selection rules according to the WF symmetry that suppress a number of transitions, as indicated by dashed arrows. The NEGF absorption coefficient which is related to the LDOS in Fig. \ref{fig:fig_4}(c) reproduces the FGR result, up to the effects of finite broadening, but including the selection rules, which is a highly non-trivial result, as it requires the consideration of the full non-locality in the evaluation of the joint density of states encoded in the polarization function \eqref{eq:polfun} \cite{ae:prb89_14}. However, the NEGF absorption coefficient based on the LDOS in Fig. \ref{fig:fig_4}(d) bares little resemblance to the square well flat band absorption, as the strong field leads to a pronounced red shift of the absorption edge and to a relaxation of selection rules due to symmetry breaking, while the unbinding of states removes the 2D character of the (joint) density of states.  

In the global detailed balance and the semiclassical picture, the absorption coefficient is the main ingredient for the computation of the current-voltage characteristics of the device at the radiative limit. In the former case, it is used for the computation of the absorptance in Eqs. \eqref{eq:phiabs} and \eqref{eq:phiem}, while in the latter it provides the generation and recombination rates via Eqs. \eqref{eq:genrate} and \eqref{eq:recrate}, respectively. In Fig. \ref{fig:fig_5}(b), the characteristics are shown for monochromatic illumination with $E_{\gamma}=1.3$  eV and at an intensity of 0.1 kW/m$^{2}$ for different theoretical descriptions: the global detailed balance picture (DB) with absorptance from Lambert-Beer law (LB) or the transfer-matrix method (TMM) for square well absorption coefficient ($\alpha_{\textrm{SW}}$); the semiclassical drift-diffusion-Poisson model (DD) coupled to TMM; and the full NEGF-Poisson model. The LB and TMM pictures provide slightly different absorptance levels, and the DD model exhibits an increased dark saturation current due to the emission into the full solid angle as compared to the loss cone in the case of the global models, and the absence of photon  recycling. Remarkably, consideration of the absorption coefficient $\alpha_{\textrm{NEGF}}$ as provided by the NEGF formalism results in semiclassical characteristics that match closely those of the full NEGF model. This justifies {\it a posteriori} the assumption of unit escape probability of carriers generated in quantum well states that is used in the semiclassical model and which is based on experimental observations for device operation at room temperature and moderate well depth \cite{nelson:93}. As in the case of the ultrathin absorber, this unit extraction efficiency is here a consequence of the long radiative lifetime, such that the escape is much faster than the recombination. In reality, escape competes with much faster non-radiative recombination channels, which in deep wells and at large forward bias corresponding to low fields results in incomplete carrier extraction. Furthermore, in a proper treatment of the charge density  component related to confined states, accumulation of charge in the QW will have effects on both the electrostatic potential as well as the recombination rate. If the confinement is neglected, as in the semiclassical approach used here, carriers are generated in extended states only and drift or diffuse away quickly. 

\subsection{Quantum dot solar cells}
In quantum dots, the deviation from bulk material properties is maximum due to confinement induced by inhomogeneity in all three spatial dimensions. There is a large  variety of quantum dot architectures studied for photovoltaic applications, ranging from hybrid organic-inorganic bulk heterojunction type solar cells based on colloidal nanoparticle to superlattices of epitaxially grown QD of regular shape \cite{nozik:02,kamat:08,nozik:10}. Here, the focus is set on regimented arrays of inorganic low band gap nanoparticles embedded in a wide band gap host material, which enable the formation of extended states due to strong coupling of adjacent nanoparticles and have been investigated for applications as tunable band gap absorbers in tandem configuration \cite{jiang:06,conibeer:08_tandem} or for intermediate band solar cells \cite{marti:01_ted}. 

\begin{figure}[b]
		\begin{center}
		\includegraphics[width=1\textwidth]{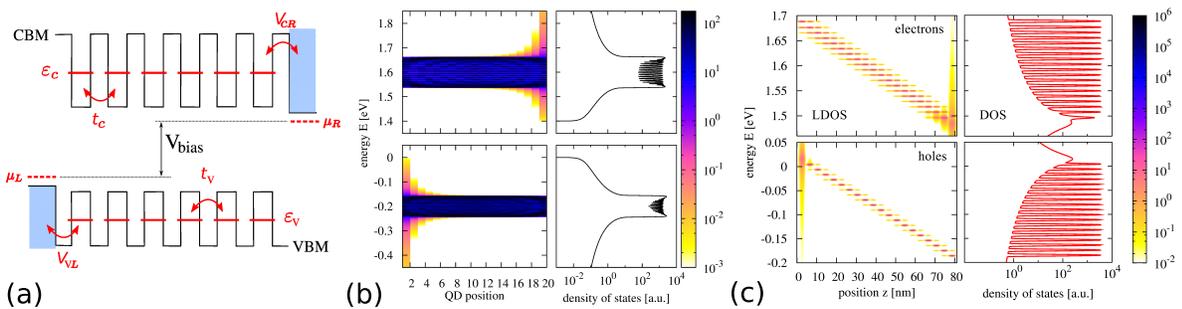}
	\caption{ (a) Schematic representation of a mesoscopic model for quantum dot array solar cells, including on-site energy, interdot hopping and coupling to carrier selective contacts \cite{berbezier:15}. (b) Local density of states (LDOS) of the lowest electron and hole minibands for a 20 QD array at vanishing field, where the states extend over the entire array. Hybridization with the bulk electrode induces strong broadening of the LDOS at the contacts  \cite{berbezier:15}. (c) In the presence of finite built-in fields, the minibands break up and the wave functions localize over a few neighboring QD. This has potentially detrimental impact on the carrier extraction efficiency, as sequential relaxation processes are required for transport to the contacts \cite{ae:oqel_12}. \label{fig:fig_6}}
		\end{center}
\end{figure}

\begin{figure}[b]
	\begin{center}
		\includegraphics[width=1\textwidth]{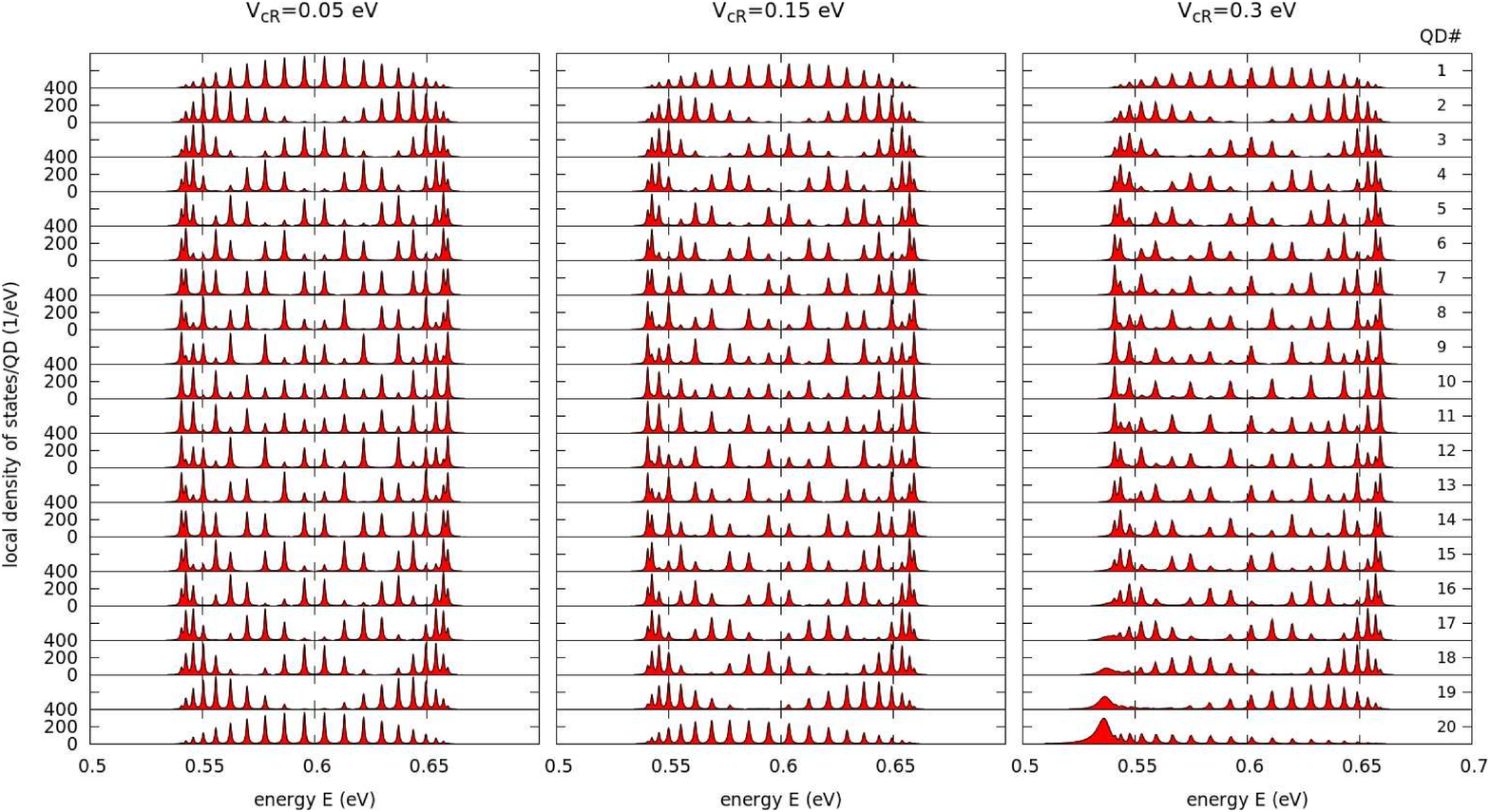}
		\caption{ Local density of states of the lowest electron “miniband” of a 20 QD array with selective contact to the dot number 20. Even at low dot-contact coupling, the finite size of the array results in a spatial variation of the site-resolved LDOS. Increasing dot-contact coupling induces a renormalization of the states adjacent to the contact, which results in a shift to lower energies and formation of a strongly localized surface state  \cite{berbezier:15}.  \label{fig:fig_7}}
	\end{center}
\end{figure}

\begin{figure}[b]
	\begin{center}
		\includegraphics[width=1\textwidth]{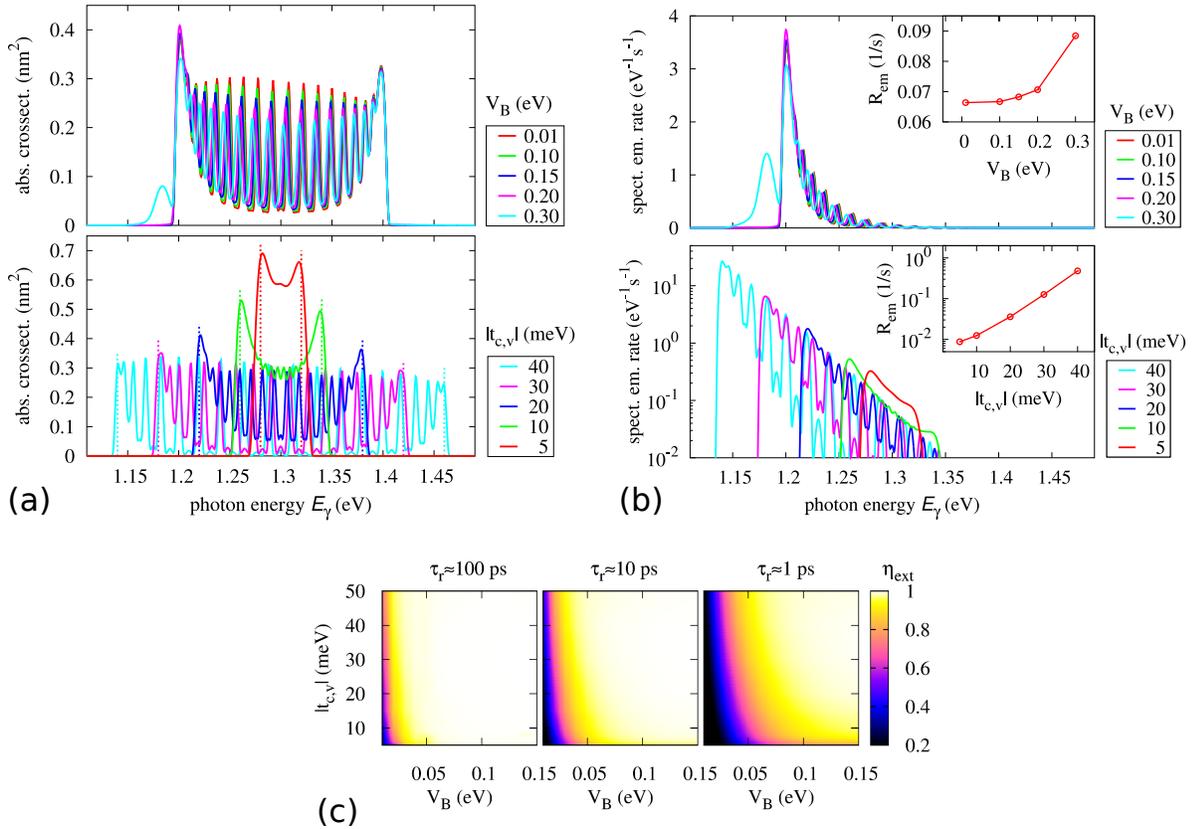}
		\caption{(a) Absorption cross section and (b) spectral emission rate for a 20 QD array with varying values of inter-dot coupling $t_{c}=t_{v}$ and dot-contact coupling $V_{B}$. The contact contributes primarily a broadening up to very strong coupling, where a localized surface state is induced. The  interdot coupling affects the bandwidth of the spectral response, including the size effective gap which has a strong impact on the magnitude of the emission. (c) Photocarrier extraction efficiency in a 20 QD array as a function of configurational parameters (inter-dot coupling $t_{c,v}$ and dot-contact coupling $V_{B}$) and carrier recombination lifetime $\tau_{r}$. The dot-contact coupling limits the extraction efficiency, while inter-dot coupling becomes critical only for very low lifetime \cite{berbezier:15}.  \label{fig:fig_8}}
	\end{center}
\end{figure}

\noindent Figure \ref{fig:fig_6}(a) shows a schematic representation of an effective mesoscopic model of such an array, furnished with on-site energies, inter-dot hopping constants and a term describing the coupling to carrier-selective contacts \cite{berbezier:15}. Contact selectivity is essential to ensure charge separation close to flat band conditions. Since atomistic resolution is out of discussion for extended arrays, a coarse grained localized basis approach is chosen where the QD wave functions are expressed in terms of QD orbitals. This amounts to field operators of the form
\begin{align}
\hat{\Psi}(\mathbf{r},t)=\sum_{i,n}\psi_{in}(\mathbf{r})\hat{d}_{in}(t),
\end{align}
with $\psi$ the QD orbital wave function and $\hat{d}$ the annihilation operator for electrons in QD orbitals. In this basis, the model corresponds to a simple nearest-neighbor tight-binding chain expressed by the Hamiltonian
\begin{align}
\hat{\mathcal{H}}_{0}^{b}=&\sum_{i=1}^{N_{QD}-1}
t_{b,ii+1}\Big[\hat{d}_{b,i+1}^{\dagger}\hat{d}_{b,i}+h.c.\Big]+\sum_{i=1}^{N_{QD}}\varepsilon_{b,i}\hat{n}_{b,i},
\end{align}
where $N_{QD}$ is the number of dots, $t_{b}$ is the inter-dot coupling, $\hat{n}_{b}\equiv\hat{d}_{b}^{\dagger}\hat{d}_{b}$ is the carrier density operator and $\varepsilon_{b}$ is the QD energy level. The dot-contact coupling is considered in the usual way by means of a contact self-energy that vanishes for closed contacts. In the flat band case displayed in Fig. \ref{fig:fig_6}(b), the coupling of the dots  leads to the delocalization of the wave functions over the entire array. At the contacts, the hybridization of the dot states with the electrode induces a broadening of the LDOS, corresponding to the reduction of lifetime due to carrier escape to the contacts. In the presence of strong built-in fields, the wave functions localize over a limited number of neighboring quantum dots, which slows down carrier extraction due to the requirement of sequential carrier relaxation processes \cite{ae:oqel_12}. This is shown in Fig. \ref{fig:fig_6}(c) for the case of a hetero-superlattice of silicon quantum dots laterally confined in silicon oxide and vertically separated by barrier layers of silicon carbide \cite{ding:energy_procedia11}. Even at flat band conditions, the site-resolved LDOS is not uniform throughout the finite-size array, as shown in Fig. \ref{fig:fig_7} for the lowest electron miniband in a 20 QD array with contact to dot number 20. For strong dot-contact coupling $V_{\mathrm{cR}}$, the states in the vicinity of the contact are renormalized, which amounts to a shift to lower energies and results in the formation of a strongly localized surface state.

The impact of inter-dot and dot contact coupling parameters $t_{\mathrm{c,v}}$ and $V_{\mathrm{B}}$ on the radiative rates is shown in Figs.~\ref{fig:fig_8}(a) and (b) displaying the absorption cross section and the spectral emission rate  of the 20 QD array for different values of the coupling parameters. Up to the critical contact coupling where surface states are formed, the main impact of the contact is to induce a slight broadening to the contributions of the individual dots to the spectra. The interdot coupling, on the other hand, determines the band width of the spectral response, which in turn has a pronounced effect on the magnitude of the emission via the size of the effective band gap. In order to study the impact of configurational parameters on the efficiency of photocarrier extraction, the ratio $\eta_{\textrm{ext}}=J_{\textrm{SC}}/J_{\textrm{abs}}$ of short circuit current to generation current as obtained from the absorption cross section is evaluated for different values $\tau_{\textrm{r}}$ of radiative lifetime and coupling parameters, respectively. As shown in Fig. \ref{fig:fig_8} for a 20 QD array at flat band conditions and under monochromatic resonant excitation with the central frequency of the miniband, it is the contact coupling  that limits the performance in all situations, while the interdot coupling becomes critical only at very low photocarrier lifetime \cite{berbezier:15}.

\section{Summary and conclusions}
The operation of nanostructure-based photovoltaic devices exhibits substantial deviations from bulk physics, not only regarding the electronic, optical and vibrational structure, but also in the dynamical processes involving nanostructure states. As a consequence, the device characteristics are determined by the operating conditions and inhomogeneities in structure and composition as well as finite size and surface effects rather than by the flat-band bulk properties of the constituent materials. The semiclassical macroscopic bulk picture conventionally used for the simulation of photovoltaic device characteristics fails to capture many of these effects by default. On the other hand, a microscopic generalization of the steady-state balance equations on the level of quantum statistical mechanics is able to provide a comprehensive quantum theory of photovoltaics at the nanoscale. This allows to assess on physical grounds the impact of configurational parameters on the photovoltaic device performance. 
Comparison of the predictions based on the different simulation approaches shows that in situations where transport is mediated by extended bulk states, consideration of the actual density of states at the operating point for the evaluation of the optical rates used in the semiclassical balance equations is sufficient to obtain the correct characteristics. The effects of potential barriers on charge extraction, on the other hand, still requires the full evaluation of the quantum transport problem. In general, a comprehensive and accurate, but still efficient assessment of nanostructure solar cell devices based on simulation will require a multiscale simulation framework relying on suitable combinations of the modelling approaches discussed in the present work.
At this point it is indicated to point out that non-classical regions in solar cells are not restricted to nano-structure based devices, but appear in the form of interface and  hetero-contact regions in a large variety of applications, from the a-Si:H/c-Si interface of the silicon heterojunction solar cell to grain boundaries and selective contacts in perovskite devices. The advanced theoretical analysis proposed here has thus applications in a wide range of current photovoltaic device research activities.

\section*{Acknowledgment}
This work has benefited from fruitful discussions within  COST action MP1406 -- MultiscaleSolar. 

\section*{References}

\bibliographystyle{ieeetr}

\begin{thebibliography}{10}

\bibitem{barnham:90}
K.~Barnham and G.~Duggan, ``A new approach to high-efficiency multi-band-gap
  solar cells,'' {\em J. Appl. Phys.}, vol.~67, pp.~3490--3493, 1990.

\bibitem{green:01}
M.~A. Green, ``Third generation photovoltaics: Ultra-high conversion efficiency
  at low cost,'' {\em Prog. Photovolt: Res. Appl.}, vol.~9, p.~123, 2001.

\bibitem{green:00}
M.~A. Green, ``Potential for low dimensional structures in photovoltaics,''
  {\em Mater. Sci. Eng., B}, vol.~74, no.~1-3, pp.~118 -- 124, 2000.

\bibitem{marti:01_ted}
A.~Marti, L.~Cuadra, and A.~Luque, ``Partial filling of a quantum dot
  intermediate band for solar cells,'' {\em IEEE Trans. Electron Devices},
  vol.~48, no.~10, pp.~2394--2399, 2001.

\bibitem{conibeer:08}
G.~Conibeer, D.~K\"onig, M.~Green, and J.~Guillemoles, ``Slowing of carrier
  cooling in hot carrier solar cells,'' {\em Thin Solid Films}, vol.~516,
  no.~20, pp.~6948 -- 6953, 2008.

\bibitem{ellingson:05}
R.~J. Ellingson, M.~C. Beard, J.~C. Johnson, P.~Yu, O.~I. Micic, A.~J. Nozik,
  A.~Shabaev, and A.~L. Efros, ``{Highly efficient multiple exciton generation
  in colloidal PbSe and PbS quantum dots.},'' {\em Nano lett.}, vol.~5, no.~5,
  pp.~865--71, 2005.

\bibitem{atwater:10}
H.~A. Atwater and A.~Polman, ``Plasmonics for improved photovoltaic devices,''
  {\em Nat. Mater.}, vol.~9, no.~3, pp.~205--213, 2010.

\bibitem{mokkapati:12}
S.~Mokkapati and K.~R. Catchpole, ``Nanophotonic light trapping in solar
  cells,'' {\em J. Appl. Phys.}, vol.~112, no.~10, p.~101101, 2012.

\bibitem{adams:10}
J.~G.~J. Adams, W.~Elder, G.~Hill, J.~S. Roberts, K.~W.~J. Barnham, and N.~J.
  Ekins-Daukes, ``Higher limiting efficiencies for nanostructured solar
  cells,'' in {\em Proceedings of SPIE}, vol.~7597, 2010.

\bibitem{ae:jstqe_13}
U.~Aeberhard, ``Simulation of nanostructure-based high-efficiency solar cells:
  Challenges, existing approaches, and future directions,'' {\em IEEE J. Sel.
  Top. Quantum Electron.}, vol.~19, no.~5, pp.~4000411--4000411, 2013.

\bibitem{araujo:94}
G.~Ara\'ujo, A.~Mart\'i, F.~Ragay, and J.~Wolter, ``Efficiency of multiple
  quantum well solar cells,'' in {\em Proc. 12th European Photovoltaic Solar
  Energy Conference}, p.~1481, 1994.

\bibitem{wuerfel:82}
P.~W\"urfel, ``The chemical potential of radiation,'' {\em J. Phys. C: Solid
  State Phys.}, vol.~15, p.~3967, 1982.

\bibitem{rau:07}
U.~Rau, ``Reciprocity relation between photovoltaic quantum efficiency and
  electroluminescent emission of solar cells,'' {\em Phys. Rev. B}, vol.~76,
  no.~8, p.~085303, 2007.

\bibitem{henrickson:02}
L.~E. Henrickson, ``Nonequilibrium photocurrent modeling in resonant tunneling
  photodetectors,'' {\em J. Appl. Phys}, vol.~91, p.~6273, 2002.

\bibitem{naser:07}
M.~A. Naser, M.~J. Deen, and D.~A. Thompson, ``Spectral function and
  responsivity of resonant tunneling and superlattice quantum dot infrared
  photodetectors using green's function,'' {\em J. Appl. Phys.}, vol.~102,
  no.~8, p.~083108, 2007.

\bibitem{pereira:98}
M.~F. Pereira and K.~Henneberger, ``Microscopic theory for the influence of
  {Coulomb} correlations in the light-emission properties of semiconductor
  quantum wells,'' {\em Phys. Rev. B}, vol.~58, pp.~2064--2076, 1998.

\bibitem{lee:prb_02}
S.-C. Lee and A.~Wacker, ``{Nonequilibrium Green's function theory for
  transport and gain properties of quantum cascade structures},'' {\em Phys.
  Rev. B}, vol.~66, p.~245314, 2002.

\bibitem{kubis:09}
T.~Kubis, C.~Yeh, P.~Vogl, A.~Benz, G.~Fasching, and C.~Deutsch, ``Theory of
  nonequilibrium quantum transport and energy dissipation in terahertz quantum
  cascade lasers,'' {\em Phys. Rev. B}, vol.~79, p.~195323, 2009.

\bibitem{steiger:iwce_09}
S.~Steiger, R.~G. Veprek, and B.~Witzigmann, ``{E}lectroluminescence from a
  quantum-well {LED} using {NEGF},'' in {\em Proceedings - 2009 13th
  International Workshop on Computational Electronics, IWCE 2009}, 2009.

\bibitem{stewart:05}
D.~A. Stewart and F.~Leonard, ``Energy conversion efficiency in nanotube
  optoelectronics,'' {\em Nano Lett.}, vol.~5, p.~219, 2005.

\bibitem{ae:prb_08}
U.~Aeberhard and R.~H. Morf, ``Microscopic nonequilibrium theory of quantum
  well solar cells,'' {\em Phys. Rev. B}, vol.~77, p.~125343, 2008.

\bibitem{ae:nrl_11}
U.~Aeberhard, ``Theory and simulation of photogeneration and transport in
  {S}i-{S}i{O}x superlattice absorbers,'' {\em Nanoscale Res. Lett.}, vol.~6,
  p.~242, 2011.

\bibitem{buin:13}
A.~Buin, A.~Verma, and S.~Saini, ``{Optoelectronic response calculations in the
  framework of k.p coupled to non-equilibrium Green's functions for
  one-dimensional systems in the ballistic limit},'' {\em J. Appl. Phys.},
  vol.~114, no.~3, p.~033111, 2013.

\bibitem{ae:oqel_12}
U.~Aeberhard, ``Effective microscopic theory of quantum dot superlattice solar
  cells,'' {\em Opt. Quantum. Electron.}, vol.~44, pp.~133--140, 2012.

\bibitem{berbezier:15}
A.~Berbezier and U.~Aeberhard, ``Impact of nanostructure configuration on the
  photovoltaic performance of quantum-dot arrays,'' {\em Phys. Rev. Applied},
  vol.~4, p.~044008, 2015.

\bibitem{cavassilas:15}
N.~Cavassilas, C.~Gelly, F.~Michelini, and M.~Bescond, ``{Reflective Barrier
  Optimization in Ultrathin Single-Junction GaAs Solar Cell},'' {\em IEEE J.
  Photovolt.}, vol.~5, no.~6, pp.~1621--1625, 2015.

\bibitem{ae:jpv_16}
U.~Aeberhard, ``Simulation of ultrathin solar cells beyond the limits of the
  semiclassical bulk picture,'' {\em IEEE J. Photovolt.}, vol.~6, no.~3,
  pp.~654--660, 2016.

\bibitem{ae:prb87_13}
U.~Aeberhard, ``{Theoretical investigation of direct and phonon-assisted
  tunneling currents in InAlGaAs/InGaAs bulk and quantum-well interband tunnel
  junctions for multijunction solar cells},'' {\em Phys. Rev. B}, vol.~87,
  p.~081302, 2013.

\bibitem{ae:jcel_11}
U.~Aeberhard, ``{Theory and simulation of quantum photovoltaic devices based on
  the non-equilibrium Green’s function formalism},'' {\em J. Comput.
  Electron.}, vol.~10, pp.~394--413, 2011.

\bibitem{kadanoff:62}
L.~P. Kadanoff and G.~Baym, {\em Quantum Statistical Mechanics}.
\newblock Benjamin, Reading, Mass., 1962.

\bibitem{keldysh:65}
L.~Keldysh, ``Diagram technique for nonequilibrium processes,'' {\em Sov. Phys.
  JETP}, vol.~20, pp.~1018--1026, 1965.

\bibitem{wang:13}
Z.~Wang, T.~White, and K.~Catchpole, ``Plasmonic near-field enhancement for
  planar ultra-thin photovoltaics,'' {\em IEEE Photon. J.}, vol.~5, no.~5,
  pp.~8400608--8400608, 2013.

\bibitem{llorens:14}
J.~M. Llorens, J.~Buencuerpo, and P.~A. Postigo, ``Absorption features of the
  zero frequency mode in an ultra-thin slab,'' {\em Appl. Phys. Lett.},
  vol.~105, no.~23, p.~231115, 2014.

\bibitem{massiot:12}
I.~Massiot, C.~Colin, N.~P\'{e}r\'{e}-Laperne, P.~Roca~i Cabarrocas, C.~Sauvan,
  P.~Lalanne, J.-L. Pelouard, and S.~Collin, ``Nanopatterned front contact for
  broadband absorption in ultra-thin amorphous silicon solar cells,'' {\em
  Appl. Phys. Lett.}, vol.~101, no.~16, p.~163901, 2012.

\bibitem{massiot:13}
I.~Massiot, C.~Colin, C.~Sauvan, P.~Lalanne, P.~R. i~Cabarrocas, J.-L.
  Pelouard, and S.~Collin, ``Multi-resonant absorption in ultra-thin silicon
  solar cells with metallic nanowires,'' {\em Opt. Express}, vol.~21, no.~S3,
  pp.~A372--A381, 2013.

\bibitem{massiot:14}
I.~Massiot, N.~Vandamme, N.~Bardou, C.~Dupuis, A.~Lemaitre, J.-F. Guillemoles,
  and S.~Collin, ``{Metal Nanogrid for Broadband Multiresonant Light-Harvesting
  in Ultrathin GaAs Layers},'' {\em ACS Photonics}, vol.~1, no.~9,
  pp.~878--884, 2014.

\bibitem{wang:13_jpv}
X.~Wang, M.~Khan, J.~Gray, M.~Alam, and M.~Lundstrom, ``{Design of GaAs Solar
  Cells Operating Close to the Shockley-Queisser Limit},'' {\em IEEE J.
  Photovolt.}, vol.~3, no.~2, pp.~737--744, 2013.

\bibitem{yang:14_jap}
W.~Yang, J.~Becker, S.~Liu, Y.-S. Kuo, J.-J. Li, B.~Landini, K.~Campman, and
  Y.-H. Zhang, ``{Ultra-thin GaAs single-junction solar cells integrated with a
  reflective back scattering layer},'' {\em J. Appl. Phys.}, vol.~115, no.~20,
  p.~203105, 2014.

\bibitem{vandamme:15}
N.~Vandamme, C.~Hung-Ling, A.~Gaucher, B.~Behaghel, A.~Lemaitre, A.~Cattoni,
  C.~Dupuis, N.~Bardou, J.-F. Guillemoles, and S.~Collin, ``{Ultrathin GaAs
  Solar Cells With a Silver Back Mirror},'' {\em IEEE J. Photovolt.}, vol.~5,
  no.~2, pp.~565--570, 2015.

\bibitem{hirst:16}
L.~C. Hirst, M.~K. Yakes, J.~H. Warner, M.~F. Bennett, K.~J. Schmieder, R.~J.
  Walters, and P.~P. Jenkins, ``{Intrinsic radiation tolerance of ultra-thin
  GaAs solar cells},'' {\em Appl. Phys. Lett.}, vol.~{109}, no.~{3}, {2016}.

\bibitem{pieters:06}
B.~Pieters, J.~Krc, and M.~Zeman, ``{Advanced numerical simulation tool for
  solar cells -- ASA5},'' in {\em Conference Record of the 2006 IEEE 4th World
  Conference on Photovoltaic Energy Conversion,}, 2006.

\bibitem{ae:17_prl}
U.~Aeberhard and U.~Rau, ``Microscopic perspective on photovoltaic reciprocity
  in ultrathin solar cells,'' {\em Phys. Rev. Lett.}, vol.~118, p.~247702,
  2017.

\bibitem{ae:apl_16}
U.~Aeberhard, ``{Impact of built-in fields and contact configuration on the
  characteristics of ultra-thin GaAs solar cells},'' {\em Appl. Phys. Lett.},
  vol.~109, no.~3, p.~033906, 2016.

\bibitem{ned:01}
N.~J. Ekins-Daukes, J.~M. Barnes, K.~W.~J. Barnham, J.~P. Connolly, M.~Mazzer,
  J.~C. Clark, R.~Grey, G.~Hill, M.~A. Pate, and J.~S. Roberts, ``Strained and
  strain-balanced quantum well devices for high-efficiency tandem solar
  cells,'' {\em Sol. Energy Mater. Sol. Cells}, vol.~68, p.~71, 2001.

\bibitem{ned:99_2}
N.~J. Ekins-Daukes, K.~W.~J. Barnham, J.~P. Connolly, J.~S. Roberts, J.~C.
  Clark, G.~Hill, and M.~Mazzer, ``{Strain-balanced GaAsP/InGaAs quantum well
  solar cells},'' {\em Appl. Phys. Lett.}, vol.~75, no.~26, pp.~4195--4197,
  1999.

\bibitem{ae:solmat_10}
U.~Aeberhard, ``Spectral properties of photogenerated carriers in quantum well
  solar cells,'' {\em Sol. Energy Mater. Sol. Cells}, vol.~94, no.~11, pp.~1897
  -- 1902, 2010.

\bibitem{ae:spie_10}
U.~Aeberhard, ``Microscopic theory and numerical simulation of quantum well
  solar cells,'' in {\em Proceedings of SPIE}, vol.~7597, p.~759702, 2010.

\bibitem{wang:12}
Y.~P. Wang, K.~Watanabe, Y.~Wen, M.~Sugiyama, and Y.~Nakano, ``{Strain-balanced
  InGaAs/GaAsP superlattice solar cell with enhanced short-circuit current and
  a minimal drop in open-circuit voltage},'' {\em Appl. Phys Express}, vol.~5,
  no.~5, pp.~1--3, 2012.

\bibitem{ae:jpe_14}
U.~Aeberhard, ``Simulation of nanostructure-based and ultra-thin film solar
  cell devices beyond the classical picture,'' {\em J. Photon. Energy}, vol.~4,
  no.~1, p.~042099, 2014.

\bibitem{ae:prb89_14}
U.~Aeberhard, ``Quantum-kinetic theory of steady-state photocurrent generation
  in thin films: Coherent versus incoherent coupling,'' {\em Phys. Rev. B},
  vol.~89, p.~115303, 2014.

\bibitem{nelson:93}
J.~Nelson, M.~Paxman, K.~W.~J. Barnham, J.~Roberts, and C.~Button, ``Steady
  state carrier escape rates from single quantum wells,'' {\em IEEE J. Quantum
  Electron.}, vol.~29, pp.~1460--1468, 1993.

\bibitem{nozik:02}
A.~Nozik, ``Quantum dot solar cells,'' {\em Physica E}, vol.~14, pp.~115 --
  120, 2002.

\bibitem{kamat:08}
P.~V. Kamat, ``Quantum dot solar cells. semiconductor nanocrystals as light
  harvesters,'' {\em J. Phys. Chem. C}, vol.~112, no.~48, pp.~18737--18753,
  2008.

\bibitem{nozik:10}
A.~J. Nozik, M.~C. Beard, J.~M. Luther, M.~Law, R.~J. Ellingson, and J.~C.
  Johnson, ``Semiconductor quantum dots and quantum dot arrays and applications
  of multiple exciton generation to third-generation photovoltaic solar
  cells,'' {\em Chem. Rev.}, vol.~110, no.~11, pp.~6873--6890, 2010.

\bibitem{jiang:06}
C.-W. Jiang and M.~A. Green, ``Silicon quantum dot superlattices: Modeling of
  energy bands, densities of states, and mobilities for silicon tandem solar
  cell applications,'' {\em J. Appl. Phys.}, vol.~99, no.~11, p.~114902, 2006.

\bibitem{conibeer:08_tandem}
G.~Conibeer, M.~Green, E.-C. Cho, D.~K\"{o}nig, Y.-h. Cho, T.~Fangsuwannarak,
  G.~Scardera, E.~Pink, Y.~Huang, T.~Puzzer, S.~Huang, D.~Song, C.~Flynn,
  S.~Park, X.~Hao, and D.~Mansfield, ``{Silicon quantum dot nanostructures for
  tandem photovoltaic cells},'' {\em Thin Solid Films}, vol.~516, no.~20,
  pp.~6748--6756, 2008.

\bibitem{ding:energy_procedia11}
K.~Ding, U.~Aeberhard, O.~Astakhov, F.~K\"ohler, W.~Beyer, F.~Finger,
  R.~Carius, and U.~Rau, ``{Silicon quantum dot formation in SiC/SiO$_{x}$
  hetero-superlattices},'' {\em Energy Procedia}, vol.~10, p.~249, 2011.

\end{thebibliography}

\end{document}